\documentclass[12pt]{article}
\usepackage{graphicx}
\usepackage[margin=1in]{geometry}
\usepackage{microtype} 

\usepackage[breaklinks, colorlinks=true, urlcolor=blue, citecolor=blue]{hyperref}

\usepackage{natbib}
\usepackage{setspace}
\usepackage{mathptmx}
\usepackage[flushleft]{threeparttablex}
\usepackage{booktabs}
\usepackage{wrapfig}
\usepackage{accessibility}
\usepackage{listings}
\usepackage{amsmath}
\usepackage[hang,flushmargin]{footmisc}
\usepackage[T1]{fontenc}
\usepackage{longtable}
\usepackage{caption}
\usepackage{booktabs}
\usepackage{xcolor}
\usepackage{silence}
\AtBeginEnvironment{thebibliography}{\singlespacing\small}
\setcitestyle{authoryear,open={(},close={)}}

\makeatletter
\newcommand\footnoteref[1]{\protected@xdef\@thefnmark{\ref{#1}}\@footnotemark}
\makeatother

\title{Are Crypto Ecosystems (De)centralizing?\\A Framework for Longitudinal Analysis}

\author
{
    \vspace{-0.5em}
    \normalsize Harang Ju, Ehsan Valavi, Madhav Kumar, Sinan Aral\\ \vspace{-0.5em}
    \normalsize Massachusetts Institute of Technology\\ \vspace{-0.5em}
    \normalsize \{harang, evalavi, madhavk, sinan\}@mit.edu
}
\date{\normalsize\today}

\begin{document} 

\maketitle 

\begin{abstract}
Blockchain technology relies on decentralization to resist faults and attacks while operating without trusted intermediaries. Although industry experts have touted decentralization as central to their promise and disruptive potential, it is still unclear whether the crypto ecosystems built around blockchains are becoming more or less decentralized over time. As crypto plays an increasing role in facilitating economic transactions and peer-to-peer interactions, measuring their decentralization becomes even more essential. We thus propose a systematic framework for measuring the decentralization of crypto ecosystems over time and compare commonly used decentralization metrics. We applied this framework to seven prominent crypto ecosystems, across five distinct subsystems and across their lifetime for over 15 years. Our analysis revealed that while crypto has largely become more decentralized over time, recent trends show a shift toward centralization in the consensus layer, NFT marketplaces, and developers. Our framework and results inform researchers, policymakers, and practitioners about the design, regulation, and implementation of crypto ecosystems and provide a systematic, replicable foundation for future studies.
\end{abstract}

\clearpage

\section{Introduction}

Human systems have historically relied on centralized entities to maintain trust \citep{ferguson2018}. But centralization creates vulnerabilities, most notably in single points of failure, susceptibility to targeted adversarial attacks, and the potential for collusion among dominant entities. In response to these challenges, crypto ecosystems, with blockchain technology as their basis, have emerged as a paradigm shift, offering a decentralized sociotechnical system to establish trust among disparate participants \citep{lamport1982byzantine, haber1991timestamp, nakamoto2008bitcoin, iansiti2017truth}. Through decentralized consensus mechanisms, crypto ostensibly mitigates the vulnerabilities of centralized systems and is resistant to a wide array of potential threats, from the double-spending of digital currencies to data corruption and collusion \citep{catalini2016simple, alzahrani2018towards, john2022bitcoin}. Thus, crypto is a foundation for alternative digital currencies, reducing settlement times and frictions in cross-border payments \citep{ammous2018bitcoin}. These innovations have led to the crypto sector experiencing substantial growth, with cryptocurrencies surpassing 3.4 trillion dollars in market capitalization in 2025 \citep{halaburda2022microeconomics}.\footnote{See \href{https://coinmarketcap.com/charts/}{coinmarketcap.com}. Accessed May 23, 2025.}

Beyond their use in digital currencies, crypto ecosystems expanded across diverse real-world applications, forming the basis of several new forms of organization, at the heart of the so-called Web3 movement. For example, they have been used to secure digital products and the ownership of non-fungible tokens, such as art and intellectual property \citep{plangger2022future, vasan2022quantifying}. Even government bonds are being integrated with crypto, and institutions like the London Stock Exchange Group are exploring crypto-powered digital markets  \citep{eib2021digitalbond, meichler2023}. Substantial investments are funding developments in new crypto applications, with prominent venture capital funds like Andreessen Horowitz having crypto funds with over \$7.6 billion in assets under management \citep{dixon2022cryptofund4}. These developments underscore the growing integration of crypto into mainstream industries and its potential to reshape economic and social infrastructures.

At the heart of crypto's promise is the principle of decentralization. This ensures that any change to a blockchain's state aligns with the consensus rules mutually agreed upon by its distributed network participants \citep{hoffman2020towards}. Thus, decentralization endows crypto with unique socioeconomic properties, including trustlessness, permissionlessness, and permanence, which, in combination, are not found in traditional systems. Decentralization also protects crypto against central points of failure and can improve welfare by mitigating information asymmetry \citep{cong2019blockchain}, facilitating efficient digital markets \citep{catalini2018antitrust}, and promoting financial inclusion \citep{cong2022inclusion}.\footnote{As an example of the dangers of centralization, in March 2022, the Ronin blockchain suffered a hack of over \$625 million, one of the largest in history, because five of the eight validators were hacked through a single vulnerability.}

While complete decentralization is impossible in public, permissionless blockchains \citep{kwon2019impossibility, bakos2021permissioned}, the promise of decentralization has spurred interest in measuring decentralization \citep{srinivasan_lee_2017, gencer2018decentralization, lin2021measuring, jia2022measuring} and its relation to other aspects of blockchains, including mining pools \citep{cong2021decentralized} and transactions scalability \citep{john2020economic, cong2022scaling}.\footnote{See Section~\ref{sec_appendix_permission} for definitions of permissionless \textit{vs} private blockchains.} However, it is yet unclear how one should measure decentralization for modern crypto ecosystems and whether they are becoming more or less decentralized over time.

Our work bridges this gap by systematically measuring the historical decentralization of \textit{crypto ecosystems}, which we define here as a set of subsystems built on and in service of a blockchain. Our contributions are two-fold. First, we developed a framework for measuring crypto decentralization across crypto ecosystems and their entire lifetime. We then compared the major metrics used to measure decentralization and outlined when certain metrics should be used. Second, we applied our framework to seven prominent crypto ecosystems (five permissionless and two permissioned) to determine whether crypto ecosystems are becoming more or less decentralized over time. We found that many crypto subsystems have become more decentralized over time. However, key subsystems, such as Bitcoin's consensus layer, NFT marketplaces, and developer subsystems, have become noticeably more centralized in recent years. In addition, we have developed a live public dashboard to report historical decentralization across crypto subsystems. These findings offer insights into the evolution of crypto decentralization, which can guide crypto researchers, operators, and policymakers in formulating strategies that foster more resilient digital economies \citep{catalini2018antitrust}.

\section{A Framework For Measuring the Decentralization of Crypto Ecosystems} \label{sec_framework}

\subsection{Decentralization across Blockchains, Subsystems, and Time}

Prior work decomposes crypto into subsystems and proposes that a blockchain is only as decentralized as its least decentralized subsystem (Table~\ref{tb_subsystems}) \cite{srinivasan_lee_2017, sai2021taxonomy}. This approach assumes that control over some large fraction of the system risks catastrophic failure. This is certainly the case for some critical subsystems, such as \textit{validators} that can double-spend and engender distrust in the entire system. However, decentralization is a multi-faceted concept that varies across different subsystems. For instance, decentralization for crypto applications fosters platform competition more than it provides security per se. Thus, decentralization has other implications in various subsystems, and it is crucial to examine each subsystem individually to get a comprehensive view of the crypto's overall decentralization.

\begin{table}[!h]
\begin{center}
\caption{Blockchain subsystems, their entities, and the contribution by which we measure decentralization.} \label{tb_subsystems}
\begin{tabular}{@{}llll@{}}
\toprule
\textit{Subsystem}      & \textit{Entity}       & \textit{Contribution} & \textit{Risks} \\
\midrule
Consensus               & Validators or miners  & Blocks    & Double-spend \\
Development             & Developer             & Commits   & Code vulnerabilities \\
Exchanges               & Exchange              & Volume    & Regulation, defaults \\
NFT marketplaces        & Marketplace           & Volume    & Smart contract vulnerabilities \\
Defi                    & Protocol              & TVL       & Smart contract vulnerabilities \\ 
\end{tabular}
\end{center}
\end{table}

In our framework of quantifying the decentralization of crypto ecosystems over time, we propose (1) a time dimension, producing panel data of crypto subsystems across time, (2) additional ecosystem layers that capture the evolution of new economically significant subsystems, and (3) Shannon entropy as a general measure of decentralization, which is critical for rigorously quantifying and facilitating comparisons across crypto ecosystems and their requisite subsystems. Thus, we obtain panel data on decentralization across blockchains and subsystems, which can reveal insights into the evolution, causes, and consequences of decentralization using techniques like causal inference \citep{ju2025decentralization}. In this framework, an \textit{entity} is any node, app, service, or individual active in a crypto subsystem, while a \textit{contribution} is the extent of an entity's participation.

We used this framework to examine the decentralization of five key blockchain subsystems, in which decentralization has varying implications (Table~\ref{tb_subsystems}): \textit{Consensus} in which validators or miners produce blocks; \textit{Development} of blockchain clients by developers; \textit{Exchanges} that allow users to buy or sell tokens; \textit{Defi protocols} which enable financial transactions, such as lending, borrowing, and trading, without traditional banking intermediaries; and \textit{NFT Marketplaces} which facilitate the buying and selling of unique, verifiable digital assets, from digital art to real estate. Examples of failures in decentralization are discussed in Section~\ref{sec_appendix_centralization}.

To provide a holistic overview, we focused our empirical analysis on a select group of seven prominent crypto ecosystems: Bitcoin, Ethereum, BNB, Solana, Tron, TON, and Ronin. These blockchains were chosen based on their market capitalization, technological significance, and diversity in use cases. At the time of this writing, Bitcoin, Ethereum, BNB, Solana, Tron, TON, and Ronin are the 1\textsuperscript{st}, 2\textsuperscript{nd}, 3\textsuperscript{rd}, 4\textsuperscript{th}, 6\textsuperscript{th}, 10\textsuperscript{th}, and 48\textsuperscript{th} Layer-1 blockchains by market capitalization.\footnote{\label{fn_marketcap}\href{https://www.coingecko.com/en/categories/layer-1/}{Coingecko}. Accessed May 29, 2025.} These include both permissionless (\textit{i.e.}, Bitcoin, Ethereum, Solana, Tron, and TON) and permissioned (\textit{i.e.}, BNB and Ronin) blockchains.\footnoteref{fn_marketcap}\footnote{While the validators can be selected by central authorities for permissioned blockchains, validator decentralization is still critical to blockchain security, even if validators for permissioned blockchains may not be politically decentralized.} 
These blockchains vary widely in their approaches, designs, and use cases (\textit{e.g.}, Ronin is primarily used for gaming), offering a diverse and generalizable view on decentralization. This depth of analysis provides a foundation for future work to expand the framework to additional ecosystems.

\subsection{Decentralization Metrics}

Central to any framework quantifying the decentralization of crypto ecosystems, are the metrics used to quantify decentralization across subsystems and time. We evaluated multiple metrics---including Shannon Entropy, the Number of Nodes, the Gini Coefficient, the Nakamoto Coefficient, the Herfindahl-Hirschman Index, and Renyi Entropy---and their unique advantages and disadvantages in particular use cases.

\subsubsection{Criteria for metrics}

To assess the effectiveness of a metric for measuring decentralization, we first establish clear criteria for what constitutes a higher or lower degree of decentralization in a system. First, decentralization, while determined by numerous qualitative factors like network resilience and lower barriers to entry, can be quantitatively summarized by two key heuristics: a system is more decentralized if it has more entities and if the contribution or control these entities exhibit in the system are more evenly distributed. While metrics do not fully address the complexities of decentralization---such as the type of consensus mechanism, the potential for collusion or Sybil attacks, or the nature of contributions---they provide a useful quantitative summary that complements sociotechnical considerations. Our analysis does not assume Sybil resistance in our data, meaning entities controlled by a single attacker are not treated as one.

Building on this foundation, we outline criteria for an ideal decentralization metric. Such a metric needs to accurately measure both the number and distribution of entities within the system. It should also be sensitive to changes within the system, capable of detecting subtle shifts in the distribution of control or influence. Furthermore, the metric must be universally applicable across different crypto subsystems, considering the entire spectrum of the system's architecture. Finally, it should be effective in making consistent comparisons across key variables across different crypto systems. These criteria ensures that the metric provides a reliable and comprehensive assessment of decentralization. In our analysis, we found that while Shannon entropy meets all of these requirements, other measures are still useful in specific scenarios.

\subsubsection{Shannon Entropy}

Mathematically, the Shannon entropy of a discrete random variable $X$ is defined as $H(X)=-\sum_{x \in X}p(x)\log_2 p(x)$. Shannon entropy captures the uncertainty in the distribution of a variable, which in the context of crypto, translates to how control or influence is spread across various participants. A higher entropy indicates a more uniform distribution of control, suggesting a more decentralized system. Conversely, a lower entropy implies the concentration of control in fewer entities, indicating a more centralized system. While prior work applied Shannon entropy to consensus layers \citep{gochhayat2020, lin2021measuring}, our work applies it longitudinally across key components of modern crypto ecosystems---developers, exchanges, DeFi, and NFT marketplaces---which are critical to understanding the full scope of crypto ecosystems.

Shannon entropy meets the established criteria for a general decentralization metric, measuring both the number and distribution of entities. Its sensitivity to small changes in system structure makes it particularly useful for tracking how decentralization evolves over time and allows for statistical inference of potential drivers of decentralization.\footnote{For systems with frequently changing distributions, such as DeFi governance and NFT marketplaces, using wider time windows can help reduce noise in entropy measurements.} A key strength of Shannon entropy is its broad applicability; it is not limited to any specific crypto system but can be used across a diverse range of systems. This wide applicability is crucial in the varied and fast-developing area of crypto. Moreover, it is a non-parametric and standardized measure, which allows researchers and practitioners to evaluate and compare decentralization in various contexts. 

\subsubsection{Number of nodes}

A naive measure of decentralization is the number of entities in a subsystem. Certainly, it is an important measure since it directly increases a blockchain's resistance to faults, attacks, and collusion: the greater the number of entities, the greater the number of entities that can become faulty and still have entities that operate the system. However, in a permissionless system, the distribution of contributions may be greatly skewed among a few entities, thus centralizing the system. To quantify the relationship between the number of nodes and entropy, we examined their correlation and found it was near-perfect for permissioned blockchains but weaker for permissionless blockchains like Bitcoin and Ethereum (see Appendix~\ref{sec_appendix_entropy_nodes}), suggesting that network growth alone does not ensure decentralization in permissionless systems \cite{kwon2019impossibility, bakos2021permissioned}.

\subsubsection{Gini Coefficient}

The Gini Coefficient is often used to measure blockchain decentralization \citep{srinivasan_lee_2017, sai2021taxonomy}. Mathematically, it is formulated as ${G=\left( \sum_{i=1}^n \sum_{j=1}^n |x_i - x_j| \right)/(2n^2\overline{x})}$, where $n$ is the number of entities and $x_i$ is the wealth, income, or other metrics of entity $i$. However, the coefficient is not a measure of decentralization but rather inequality in a distribution \citep{dixon1987bootstrapping}. 

Despite its common use in measuring decentralization, the Gini coefficient should be used where inequalities in distributions are the primary focus, rather than decentralization. In the trivial case where the population size is one, the Gini coefficient is zero, indicating no inequality. But this outcome fails to capture decentralization, as centralization is maximized with only one entity but the Gini coefficient measures this case as being characterized by complete uniformity and no inequality. Beyond the trivial case, the Gini coefficient fails to capture differences in decentralization where inequality is equivalent but decentralization is not.\footnote{For example, probability distributions of $a=\{1\}$, $b=\{0.5, 0.5\}$, and $c=\{0.25, 0.25, 0.25, 0.25\}$ all have Gini coefficients of zero, indicating perfect equality, but the distribution $c$ is more decentralized than $b$ (and $b$ more than $a$).} Even in the case without perfect equality, the Gini coefficient fails to distinguish between levels of decentralization.\footnote{Given $d=\{0.65, 0.35\}$ and $e=\{0.3, 0.3, 0.3, 0.1\}$, $G(d)=G(e)=0.15$, but $e$ is more decentralized than $d$.} However, Shannon entropy captures the differences between both the number of nodes and inequality across distributions.\footnote{The entropies of $a$, $b$, and $c$ are $0$, $1$, and $2$, respectively, reflecting the intuition for the decentralization of those distributions. Entropy captures the differences between $d$ and $e$ as well, with $H(e)\approx0.93$ and $H(d)\approx1.89$.}

\subsubsection{Nakamoto Coefficient} \label{sec_metric_nakamoto}

The Nakamoto coefficient was developed by \cite{srinivasan_lee_2017} to measure the minimum number of entities required to achieve 51\% of the contributions to a blockchain system. The Nakamoto coefficient is formulated as ${N_s=\min \{ k \in [1,...,K] : \sum_{i=1}^k p_i \ge 0.51 \}}$ over a system $s$ with $p_1>...>p_K$ as the proportions of contributions by each of the $K$ participants. The Nakamoto Coefficient is highly relevant for systems that require state synchrony based on a threshold, such as block production in blockchain consensus. In blockchain consensus, $N_s$ entities can collude to produce blocks containing arbitrary transactions and still have the longest chain of blocks, which serves as proof of the sequence of events witnessed by the participants. Thus, the Nakamoto coefficient is an interpretable and critical measure of the decentralization of the consensus layer.

The Nakamoto coefficient, however, is lacking in its ability to characterize several key aspects of crypto ecosystem decentralization. First, the Nakamoto coefficient assumes that the system under study has a threshold at which an individual or a coalition can control the entire system. This is true in blockchain consensus, which often has a threshold for achieving consensus, but not in many other crypto systems, especially as they expand beyond the consensus layer to applications and platforms. Second, the Nakamoto Coefficient does not contain information about the distribution of entities outside of $\{p_1,...,p_k\}$, which we call the \textit{Nakamoto Set}. To illustrate, we perform knockout simulations of the Nakamoto Set in Section~\ref{sec_appendix_nakamoto}.
Thus, one must take care in using the Nakamoto coefficient for non-thresholded systems or when concerned with the whole distribution of entities.

\subsubsection{Herfindahl-Hirschman Index}

The Herfindahl-Hirschman Index (HHI) is a measure of market concentration, and it is formulated as $\text{HHI}=\sum_{i=1}^N(\text{MS}_i)^2$,
where $\text{MS}_i$ is the market share of the $i^{\text{th}}$ firm or entity \citep{rhoades1993herfindahl}. Out of all of the measures examined here, HHI is the most comparable to entropy, as they are inversely proportional to each other. However, HHI is range-bound between 0 and 1 and thus does not scale with the number of entities. As the number of entities increases, HHI becomes infinitesimally smaller, making it difficult to compare between highly decentralized systems. In contrast, entropy scales logarithmically with the number of entities and is thus more intuitive and conducive to statistical inference.

\subsubsection{Renyi Entropy}

Renyi entropy is a generalization of Shannon entropy \citep{renyi1961entropy} and is defined for a set of probabilities $p=(p_1,...,p_n)$ and a non-negative parameter $\alpha$ not equal to 1 as $H_{\alpha}(p) = \frac{1}{1-\alpha} \log \left( \sum_{i=1}^{n} p_i^{\alpha} \right)$. This parameterized entropy allows for a nuanced analysis of distributions, offering a spectrum of diversity indices based on the value of $\alpha$. As $\alpha$ approaches 1, Renyi entropy converges to Shannon entropy, and by adjusting $\alpha$, one can give more or less weight to parts of the distribution that are of particular interest, making it a versatile and encompassing tool for diversity assessment. See further discussion in Section~\ref{sec_appendix_renyi}.

\section{Results}

Having developed and justified a systematic framework for longitudinal analysis of crypto decentralization, we applied this framework to five subsystems (Table~\ref{tb_subsystems}) of prominent blockchains to examine whether crypto has become more or less decentralized over time. We use Shannon entropy in this section. For readers interested in a dynamic view of our findings, we have developed a live dashboard at the \href{https://deepnote.com/app/harang/Crypto-Decentralization-Dashboard-905957ed-ef82-4fc5-b018-d0b021ade927}{\textit{Crypto Decentralization Dashboard}}.

\subsection{Consensus} 

\begin{figure}[ht]
  \centering
  \includegraphics[width=\linewidth]{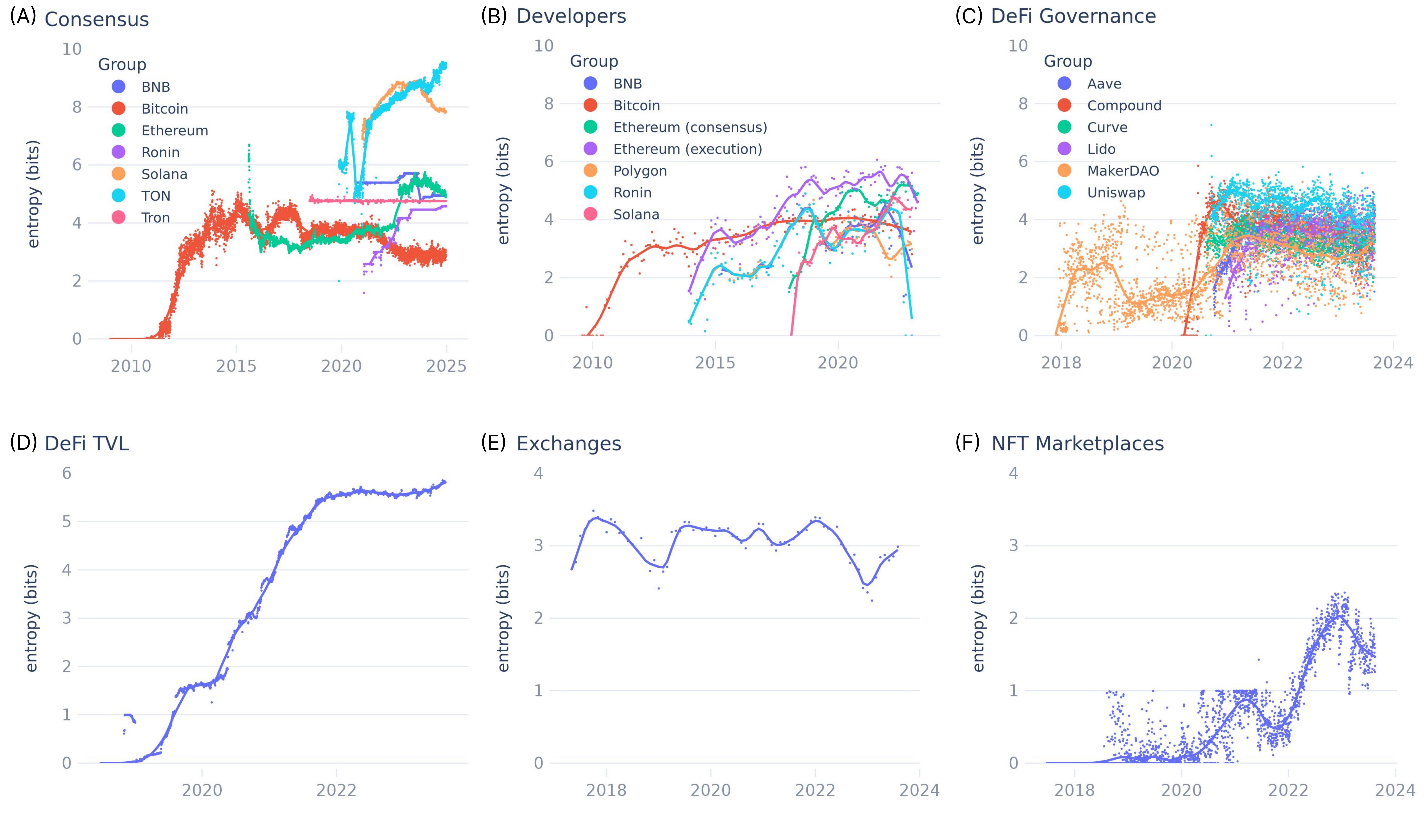}
  \caption{Daily entropy across crypto ecosystems.}
  \label{fig_entropy}
\end{figure}

The consensus layer is fundamental to a blockchain's architecture. It ensures network-wide agreement on the blockchain's state, such as essential data like token balances, by verifying the legitimacy of blockchain transactions. Our measurement of daily entropy data reveals that most blockchains have trended towards decentralization over time (Figure~\ref{fig_entropy}A). Permissioned blockchains like Ronin and BNB have exhibited stepwise entropy increases, indicative of centralized decisions to expand the validator count. Similarly, the permissionless blockchains Ethereum and Solana have seen increased decentralization; however, Bitcoin's decentralization has diminished since late 2021, which suggests a potential relative risk to Bitcoin's integrity.

\subsection{Client Development}

\begin{figure}[ht]
  \centering
  \includegraphics[width=\linewidth]{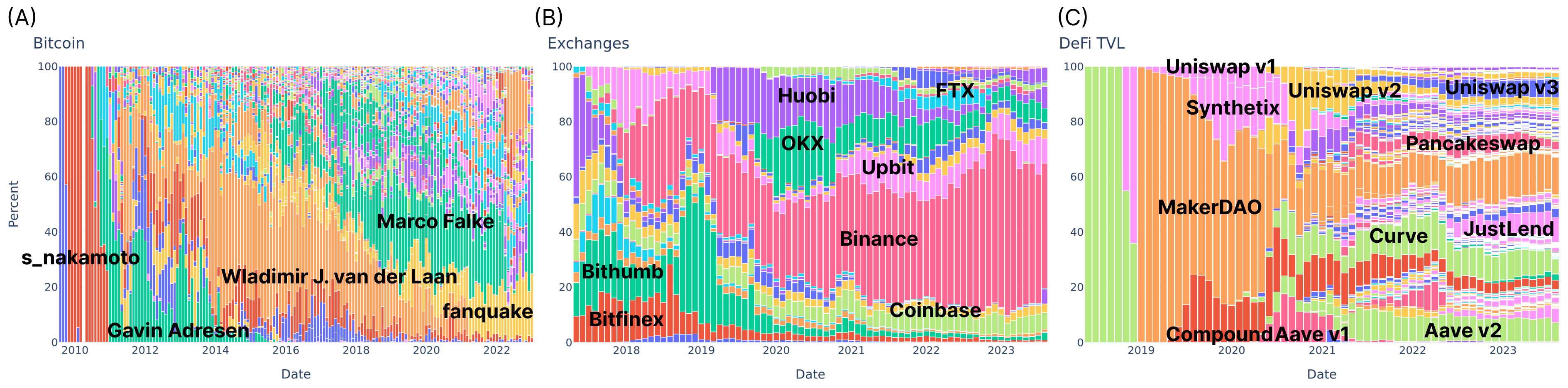}
  \caption{Examples of labeled distributions of contributions by entities: (A) monthly commits by Bitcoin developers, (B) monthly volume across centralized exchanges, and (C) daily TVL in Defi protocols.}
  \label{fig_distributions}
\end{figure}

Blockchain clients are the software that powers servers on blockchain networks and are instrumental to the functionality of blockchains. The decentralization of the development of clients is thus pivotal for ensuring blockchain integrity by preventing or fixing code vulnerabilities in the client software.\footnote{As a real-world example, on June 22, 2018, a few Bitcoin developers discreetly identified and patched a vulnerability with the potential to incapacitate the entire network \citep{fuller2020bitcoin}.} To measure decentralization in blockchain client development, we obtained the monthly number of commits created by each developer for all clients of a blockchain and calculated the monthly entropy for each blockchain. We found that development decentralization has generally been on the rise across all blockchains (Figure~\ref{fig_entropy}B). This general trend toward decentralization is accompanied by more diverse groups of developers across time, as we can see in monthly distributions of git commits for Bitcoin Core in Figure~\ref{fig_distributions}A. However, we observed downtrends in development decentralization since mid-2020 for Bitcoin and since 2022 for the other blockchains, especially in the permissioned blockchains BNB and Ronin.\footnote{Ronin's \href{https://github.com/axieinfinity/ronin/commits/master}{development} has been primarily done by a single developer since mid-2022. Accessed September 7, 2023.} The stark divergence between these and their open-source counterparts is noteworthy and reflects findings by \cite{lakhani2003open} on motivations for developers who contribute to open-source software.

\subsection{Exchanges}

Centralized exchanges, or simply \textit{exchanges}, are integral to crypto ecosystems, as they enable the trading of tokens and are the gateways for transitions between tokens and fiat currency. A scenario where only one centralized exchange exists would pose significant risks: if that exchange were to fail or become unavailable, users would be left without the ability to exchange tokens for fiat currency or vice versa. To quantify the decentralization of exchanges, we measured the entropy of their monthly volume (Figure~\ref{fig_entropy}E). In this context, higher entropy indicates a more even distribution of trading volume across multiple exchanges. Unlike the decentralization of other subsystems, we observed that the decentralization of exchanges has remained range-bound between 2.24 and 3.48 bits, despite the introduction of new exchanges over time (Figure~\ref{fig_distributions}B). This limited range of decentralization suggests that both drivers of centralization, such as network effects, and decentralization, such as lower barriers to entry, may be balanced in the market of exchanges.

\subsection{Defi Protocols (TVL) and Governance}

Decentralized finance, or \textit{Defi}, represents a suite of applications run on blockchains that enable users to trade, lend, and borrow tokens. Defi protocols hold an impressive \$138 billion in value, termed as total value locked (TVL), but they are susceptible to vulnerabilities, which has resulted in \$9 billion lost to hacks.\footnote{See the \href{https://defillama.com/}{Defi Llama dashboard} and \href{https://defillama.com/hacks}{hacks on DefiLlama}. Accessed December 19, 2024.} While such hacks do not result in a catastrophic failure of the entire system as in the consensus layer, increasing centralization in Defi protocols heightens the risk of large-scale hacks. Our analysis evaluates the decentralization of Defi protocols using the entropy of daily TVL across Defi protocols (Figure~\ref{fig_entropy}D). We observed an \textit{S curve} in the decentralization of Defi protocols from 2018 to 2023, in which there was a phase of initial slow growth, followed by a rapid increase in decentralization as more players entered the market, and finally a plateau since late 2021 as the market matures and growth stabilizes with a few dominant protocols (Figure~\ref{fig_distributions}C).

Within each Defi protocol, governance plays a pivotal role in steering the direction of the platform and making decisions related to upgrades, fee structures, and other operational parameters. Governance typically operates on a decentralized model where token holders have voting rights proportional to their holdings \citep{cong2022inclusion}. Our analysis evaluates the decentralization of Defi governance by assessing the entropy of governance token distributions among wallets (Figure~\ref{fig_entropy}C). Notably, the protocols quickly stabilized between 3 and 4.5 bits (Figure~\ref{fig_entropy}E) despite differences in the initial and ongoing distributions of tokens across protocols, such as airdropping to early users, distributing to founders and investors, or rewarding liquidity providers or validators. This convergence suggests that token distribution strategies are limited in their ability to meaningful change long-term decentralization, shown theoretically by \cite{bakos2022blockchains}.

\subsection{NFT Marketplaces}

Marketplaces for non-fungible tokens (NFTs) have emerged as a major category of crypto applications. NFTs are unique digital assets on the blockchain that have captured significant attention for their role in digitizing art, collectibles, and other tangible assets, ensuring their authenticity and ownership. Centralization in marketplaces could not only make artists and collectors vulnerable to potential software risks but might also lead to undue influence over NFT prices, curation, and visibility, thereby challenging the values of decentralization and open access.

Here, we quantified the decentralization of NFT marketplaces as the entropy of daily volume across NFT marketplaces (Figure~\ref{fig_entropy}F). In contrast to Defi, NFT marketplaces were quite centralized with an average entropy of 0.29 bits before 2022. An entropy below 1 essentially implies a monopoly on NFT marketplaces, and we indeed observe that for most of the time, OpenSea was the major marketplace. However, when Blur launched its marketplace beta on May 4, 2022, we can see that average entropy increased over 1 bit and reached 2.35 bits at its peak when OpenSea and Blur were competing roughly equally for market share. The dynamics observed between OpenSea and Blur highlight the complexities of market competition within decentralized ecosystems, where network effects and barriers to entry can shape the competitive landscape.\footnote{See Section~\ref{sec_appendix_search} for a discussion of why decentralization trends differed for NFT marketplaces \textit{versus} Defi.}

\section{Conclusion}

To what extent are crypto ecosystems fulfilling their promise of decentralization? To answer this question, we developed and applied a framework to measure historical trends in crypto decentralization across subsystems and time in seven prominent crypto ecosystems. We found that crypto ecosystems are largely becoming more decentralized over time, with notable exceptions like Bitcoin's consensus layer. These trends are often shaped by deliberate community efforts, such as promoting client diversity.\footnote{\href{https://clientdiversity.org/}{Client Diversity Reports}. Accessed on December 16, 2024.} Additionally, we highlighted the patterns seen in the decentralization of exchanges, Defi, and NFT marketplaces, which point to the role of market dynamics.

\clearpage
\bibliography{ref}

\clearpage
\appendix
\setcounter{page}{1}

\section*{\Large Appendix}

for \textit{Are Crypto Ecosystems (De)centralizing? A Framework for Longitudinal Analysis}
\\Harang Ju, Ehsan Valavi, Madhav Kumar, Sinan Aral

\vspace{1em}

\renewcommand{\thesubsection}{\Alph{subsection}}
\renewcommand\thefigure{\thesubsection.\arabic{figure}}
\renewcommand\thetable{\thesubsection.\arabic{table}}
\renewcommand\theequation{\thesubsection.\arabic{equation}}
\setcounter{figure}{0} 
\setcounter{table}{0} 
\setcounter{equation}{0} 

\subsection{Research Methods}

\subsubsection{Data}

We used public blockchain data from various sources for each crypto subsystem. Those subsystems and the entities that comprise each system are consensus by nodes, development by developers, exchanges by monthly volume on exchanges, decentralized finance (Defi) by total value locked (TVL), Defi protocols by governance token ownership, and non-fungible token (NFT) marketplaces by dollar volume. Conceptually, the subsystems can be grouped into the consensus layer (\textit{i.e.}, the blockchain itself), the infrastructure and software that support the consensus layer (\textit{i.e.}, exchanges and developers), and the applications that run on top of the consensus layer (\textit{i.e.}, DeFi protocols and NFTs).

For each blockchain in the \textit{Consensus} subsystem, we obtained the daily number of blocks mined by each validator (for Proof-of-Stake) or miner (for Proof-of-Work) on the data provider \href{https://dune.com/}{Dune Analytics}. We collected public data for the following blockchains: Bitcoin, Ethereum, Solana, BNB, Tron, TON, and Ronin. To maintain consistency across PoS, PoW, or other blockchain specifications and to avoid using potentially biased estimations of unobserved nodes, we used the addresses that received consensus rewards as individual entities. For Ethereum post-Merge, we accounted for the obfuscation of validator addresses with the introduction of proposal-builder separation (PBS; see Section~\ref{sec_appendix_mev}). For Bitcoin, we divided the attribution of the block proportionally among the recipients of the block reward to account for mining pools; for all others, we attributed each block to a single node.\footnote{Bitcoin initially did not have ``addresses'' as Bitcoin and other blockchains do today. Instead, Bitcoin used the raw public key, denoted as \textit{pubkey}. For the purposes of measurement, we mark the addresses of all pubkey transactions as \textit{Unknown}, thus underestimating the decentralization of Bitcoin for the first one to two years. See \citep{hong2023analyze} for a detailed explanation of Bitcoin addresses.} Thus, unless a mining pool directly sent mining rewards to individual participants in the coinbase transaction, all mining and staking pools were treated as individual entities, as consensus rewards are transferred through centralized services to the contributing nodes. Moreover, since blockchain addresses are not Sybil-resistant, we cannot obtain Sybil-resistant decentralization metrics, implying that the actual decentralization of the blockchain is less than or equal to the decentralization metrics obtained here.

For \textit{Defi TVL}, we used daily TVL per Defi protocol which is freely and publicly available on \href{https://defillama.com}{DefiLlama}. For \textit{Defi Governance}, we used the daily number of tokens held across wallets on \href{https://dune.com/}{Dune Analytics}. For \textit{NFT Marketplace}, we used daily marketplace volume in dollars for each NFT marketplace using the API from \href{https://reservoir.tools}{Reservoir} through the SQL interface on \href{https://dune.com/}{Dune Analytics}. For \textit{Exchanges}, we scraped data from \href{https://www.theblock.co/data/crypto-markets/spot/cryptocurrency-exchange-volume-monthly}{TheBlock} for the top 38 exchanges. For the \textit{Development} subsystem, we obtained the monthly number of commits created by each developer from the public GitHub repository of each client (\textit{i.e.}, software that connects computers to a blockchain network) and grouped them by the blockchain. For example, we grouped the multiple Ethereum Execution clients into the category \textit{ethereum-execution}. Specifically, we examined the \href{https://github.com/bitcoin/bitcoin}{Bitcoin Core} client for Bitcoin, the \href{https://github.com/anza-xyz/agave}{Agave} client for Solana, the \href{https://github.com/maticnetwork/bor}{Bor} client for Polygon, the \href{https://github.com/bnb-chain/bsc}{BSC} client for BNB, the \href{https://github.com/axieinfinity/ronin}{Ronin} client for Ronin, the \href{https://github.com/ethereum/go-ethereum}{Go-Ethereum}, \href{https://github.com/NethermindEth/nethermind}{Nethermind}, \href{https://github.com/hyperledger/besu}{Besu}, \href{https://github.com/ledgerwatch/erigon}{Erigon}, \href{https://github.com/ethereumjs/ethereumjs-monorepo}{EthereumJS}, \href{https://github.com/status-im/nimbus-eth1}{Nimbus-ETH1}, \href{https://github.com/paradigmxyz/reth}{Reth}, and \href{https://github.com/erigontech/silkworm}{Silkworm} clients for Ethereum's execution layer, and the \href{https://github.com/prysmaticlabs/prysm}{Prysm}, \href{https://github.com/sigp/lighthouse}{Lighthouse}, \href{https://github.com/status-im/nimbus-eth2}{Nimbus-ETH2}, and \href{https://github.com/Consensys/teku}{Teku} clients for Ethereum's consensus layer.

\subsubsection{Distributions}

Before measuring decentralization, we aggregated the data into daily distributions of counts of contributions across entities. We defined an \textit{entity} here as a node, an app, a service, or an individual that contributes to a crypto subsystem. The \textit{contribution} refers to the number of instances in which an entity has contributed to a blockchain subsystem. For example, for the consensus layer of a PoS blockchain like Ethereum, the \textit{entity} is a validator, and the \textit{contribution} is the number of blocks mined per entity on a particular day. For Bitcoin, we used rewards as the contribution, but for all other blockchains, we used validated blocks as the contribution. For the software developers of Ethereum, the \textit{entity} is an individual developer, and the \textit{contribution} is the number of commits each developer has produced for a particular month. Table~\ref{tb_subsystems} shows the definitions for \textit{entities} and \textit{counts} for each crypto subsystem.

\subsubsection{Identifying Ethereum Validators post Proposer-Builder Separation (PBS)} \label{sec_appendix_mev}

Proposer-builder separation (PBS) creates an open market for block building, where transaction ordering can extract maximal extractable value (MEV) \citep{daian2019mev}. Flashbots developed \href{https://github.com/flashbots/mev-boost}{MEV-Boost}, a middleware for PBS which launched on Ethereum during the Merge (block 15537940). See the \href{https://etherscan.io/block/0x7d57a1d26f71724737f5dc780ca2dfb778c2fc5be29bcaeb7b989b768953aabe}{deployment transaction} and Flashbots's \href{https://boost.flashbots.net/mev-boost-status-updates/mev-boost-status-update-sep-9-sept-22-2022}{blog post}. MEV-Boost allows validators to sell block space to builders, increasing staking rewards by over 60\%.\footnote{See \href{https://hackmd.io/@flashbots/mev-in-eth2}{hackmd.io}. Accessed April 25, 2024.}

MEV-Boost complicates identifying proposers because block rewards go to builders, not validators. To address this, we identified proposers by tracking transactions where builders transfer rewards to proposers, which verify receipt before signing the block. We used labeled MEV builder addresses from Etherscan\footnote{See \href{https://etherscan.io/accounts/label/mev-builder}{etherscan.io}. Accessed April 25, 2024.} and lists available on Dune.\footnote{See \href{https://dune.com/queries/3665816}{Dune query}.} For builders using alternate addresses, we manually labeled these addresses.\footnote{See \href{https://dune.com/queries/3669067}{Dune query}.} Some builders are also proposers and are labeled accordingly.\footnote{See \href{https://dune.com/queries/3665820}{Dune query}.} 
To compile the list of block reward recipients, we queried addresses that deposited ETH to the Beacon staking contract, top proposer fee recipients from Etherscan,\footnote{See \href{https://etherscan.io/accounts/label/proposer-fee-recipient}{etherscan.io}. Accessed April 25, 2024.} and MEV builders who received rewards from other builders.\footnote{Example address: \href{https://etherscan.io/address/0x7e2a2FA2a064F693f0a55C5639476d913Ff12D05}{0x7e2a2FA2a064F693f0a55C5639476d913Ff12D05}.} The final list is available on Dune.\footnote{See \href{https://dune.com/queries/3665820}{Dune query}.}

\subsubsection{Code and data availability}
All source code is publicly available at \href{https://deepnote.com/workspace/harang-beb46cf3-72b8-44ac-abf7-de74ba5eb655/project/MIT-IDE-Crypto-Decentralization-Dashboard-905957ed-ef82-4fc5-b018-d0b021ade927/notebook/4-analysis-10cad5391d2e4c249334cd804a101df1}{deepnote.com}. SQL queries are publicly accessible for \href{https://dune.com/queries/2448502}{the consensus layer}, \href{https://dune.com/queries/2453824}{Defi governance tokens}, and \href{https://dune.com/queries/2453864}{NFT marketplace volume}. All data sources are publicly accessible. Aggregated daily entity-level data is available on \href{https://doi.org/10.6084/m9.figshare.24588711}{Figshare}.

\subsection{Supplementary Discussions}

\subsubsection{Permissioned-\textit{vs}-Permissionless, Private-\textit{vs}-Public Blockchains} \label{sec_appendix_permission}

We define permissioned-\textit{vs}-permissionless and private-\textit{vs}-public blockchains here. In a permissionless or open blockchain, anyone can join the network as a node or validator and publish smart contracts. In a permissioned or closed blockchain, one must acquire permission from the governing body, \textit{e.g.}, firms or groups of validators, to join the network or publish smart contracts. In either type of blockchain, the transaction and block data are often publicly accessible, as they are in the permissionless and permissioned blockchains measured in this study. A private blockchain is restricted in terms of who can view the transaction and block data, limiting access to authorized participants only. These blockchains are typically used within organizations or consortia where privacy and confidentiality are essential. In contrast, a public blockchain allows anyone to read, verify, and audit the data, promoting transparency and decentralization. We do not measure any private blockchains in this study.

\subsubsection{Perils of Centralization} \label{sec_appendix_centralization}

In the Consensus layer, centralization can directly lead to double-spending through which hackers or colluders can change the balance of tokens on their account and ultimately lead to distrust and catastrophic failures in the system. A notable example of catastrophic failure (though not one caused by centralization) is the collapse of the Terra blockchain, during which almost \$45 billion of Terra Luna's market capitalization went to zero in a single week \citep{miller2022terra}. In the Development layer, centralization, as measured in software commits by different developers, can lead to the introduction of malicious code or the inability to patch software bugs. In the Exchange layer, centralization can lead to over-reliance on services that are vulnerable to regulatory action or defaults, as seen in the collapse of FTX \citep{yaffebellany2022sam}. In Defi, centralization, as measured by total value locked (TVL) or by the distribution of ownership of governance tokens, can lead to greater exposure of capital to hacks or collusion. In NFT Marketplaces, centralization, as measured by trading volume, can lead to security risk due to smart contract vulnerabilities and anti-competitive practices, where dominant platforms may stifle innovation and limit market access for new entrants. In each layer, the implications of decentralization are context-specific, underscoring the need for a more general approach to assessing crypto systems.

\subsubsection{Benefits of Renyi Entropy} \label{sec_appendix_renyi}

The benefit of Renyi entropy is its ability to emphasize different aspects of the probability distribution. For example, setting $\alpha$ greater than 1 increases the weight given to larger proportions, thus amplifying the influence of dominant entities within the blockchain. This can be useful when assessing the risk of centralization or the potential for 51\% attacks. Conversely, a smaller $\alpha$ can shed light on the contributions of smaller entities, painting a more detailed picture of the system's inclusivity and the dispersion of control or wealth among participants. The adaptability of Renyi Entropy could aid future research in which specific distributional characteristics warrant closer examination and the selection of $\alpha$ can be justified in the context of the study. Here, we report Shannon entropy for its objectivity and forward applicability.

\subsubsection{Search Costs in Fungible \textit{vs} Non-Fungible Tokens} \label{sec_appendix_search}

While NFT marketplaces and Defi protocols are similar (\textit{i.e.}, they both allow users to trade tokens), their difference in entropy suggests a fundamental difference in the platforms—the difference in search costs between fungible and non-fungible tokens. Fungible tokens, commonly used in Defi platforms, are interchangeable, allowing users to easily compare prices across platforms. This results in lower search costs and fosters a competitive environment, pushing platforms towards decentralization as users frequently switch based on preferences and offers. In contrast, non-fungible tokens (NFTs) are inherently unique digital assets, making the search and comparison process across platforms more complex and time-consuming. This intricacy often ties users to specific platforms where they find certain NFTs or have had trading success, inadvertently leading to a centralized marketplace, as evidenced by OpenSea's dominance until Blur's emergence. The distinctiveness, rarity, and sentimental value of NFTs mean their marketplaces face different competitive dynamics than those handling fungible tokens. As the industry continues to evolve, understanding these nuances becomes vital for designing platforms that balance user needs with the broader objectives of decentralization and transparency.

\subsection{Supplementary Results}

\subsubsection{Correlation between Entropy and the Number of Nodes}
\label{sec_appendix_entropy_nodes}

To quantify the relationship between the number of nodes and entropy, we examined the correlation between daily entropy and the daily number of nodes that produced blocks. In permissioned blockchains, we observed a near-perfect correlation of 0.99 ($p\ll0.001$) for Ronin, since a node can only validate if it has permission from the company operating the blockchain. Thus, all nodes participate equally in consensus with other nodes. Similarly, TRON, which uses a delegated Proof-of-Stake (DPoS) mechanism where only a limited set of Super Representatives produce blocks, exhibited a strong correlation of 0.95 ($p\ll0.001$). Although TRON is nominally permissionless, the restricted number of block producers introduces a more structured and centralized consensus model, akin to permissioned systems.

In the case of BNB, which also limits the number of validators, the correlation was lower at 0.71 ($p\ll0.001$). This discrepancy likely reflects differences in the underlying validator selection and operational dynamics. BNB employs a Proof-of-Staked-Authority (PoSA) consensus mechanism, where validator participation is influenced by staking dynamics and periodic elections. As a result, the number of active validators and their consensus power distribution can vary over time, introducing more entropy variability than Ronin’s static and tightly controlled setup.

In contrast, public blockchains like Bitcoin, Ethereum, and TON, which rely on permissionless consensus mechanisms, showed positive but less pronounced correlations. Ethereum had a correlation of 0.75 ($p\ll0.001$), Bitcoin a much lower 0.29 ($p\ll0.001$), and TON, which uses a Proof-of-Stake model with validator staking, a moderate correlation of 0.86 ($p\ll0.001$). The disparity in correlation coefficients between these blockchains can be attributed to two main reasons.

First, in permissionless blockchains, participants buy consensus power either with hardware and electricity for Proof-of-Work or with tokens for Proof-of-Stake. In contrast, permissioned blockchains often have fewer validators to which they assign equal consensus power. In TRON's case, the delegated model restricts block production to a small, elected group, which reduces variability in entropy despite its nominal permissionless nature. In BNB, the capped number of validators combined with the dynamic staking process results in a somewhat hybrid structure, leading to a weaker correlation than fully controlled permissioned systems like Ronin. Thus, in permissionless chains like Ethereum, Bitcoin, and TON, some participants purchase more power than others, leading to a skewed distribution of consensus power. Second, staking and mining pools consolidate consensus power among the few, especially for Bitcoin, in which barriers to entry render solo mining ineffective unless the miner has substantial hash power.

This disparity underscores that mere network growth does not guarantee decentralization in permissionless blockchains, as shown by \cite{kwon2019impossibility} and \cite{bakos2021permissioned}. Achieving and maintaining decentralization in permissionless blockchains may therefore necessitate specific design and operational strategies.

\subsubsection{Knockout Simulations of the Nakamoto Set} \label{sec_appendix_nakamoto}

Here, we demonstrate that the Nakamoto Coefficient does not contain information about the distribution of entities outside of $\{p_1,...,p_k\}$. First, we define a \textit{Nakamoto Set} as the set of entities $\{1,...,k\}$ that determine the Nakamoto Coefficient $N_s$ and a \textit{Non-Nakamoto Set} as the set $\{k+1,...,K\}$. When a fault or attack takes the entities in the Nakamoto Set offline, the Non-Nakamoto Set will then become the entire network. Thus, the distribution of the Non-Nakamoto Set becomes critical when the Nakamoto Set is compromised due to faults. The Nakamoto Coefficient contains no information about the Non-Nakamoto Set other than that, before the fault or attack, then it comprised less than 51\% of the network. In contrast, other measures, such as Shannon Entropy, contain information about the whole distribution. To quantify the difference between these metrics in empirical data, we performed a ``knockout'' simulation by removing the Nakamoto Set for each day of data and plotting the measures before and after the removal. We can see in Table~\ref{tb_knockout} that Pearson's correlation coefficients between pre- and post-knockout measures are higher and more significant for entropy than for the Nakamoto Coefficient, demonstrating the relative advantage of entropy in holistically measuring decentralization in real crypto systems.

\begin{table}[ht]
\centering
\caption{Pearson's correlation coefficients between daily measures before and after the knockout of the Nakamoto set.} \label{tb_knockout}
\begin{tabular}{@{\extracolsep{20pt}}lll@{}}
\toprule
& \multicolumn{2}{c}{Pearson's R} \\
Blockchain & Entropy   & Nakamoto\\
\midrule
Bitcoin             & 0.71$^{***}$	     & 0.01$^{}$\\
Ethereum            & 0.93$^{***}$	     & 0.74$^{*}$\\
Solana              & 0.97$^{***}$	     & 0.60$^{*}$\\
BNB                 & 1.00$^{***}$	     & 0.08$^{*}$\\
Tron                & 0.96$^{***}$       & -0.0028\\
TON                 & 0.98$^{***}$	     & 0.85$^{*}$\\
Ronin               & 0.98$^{***}$	     & 0.97$^{*}$\\
\end{tabular}
\begin{tablenotes}
\footnotesize
\item Note: $^{*}$p$<$0.05; $^{**}$p$<$0.01; $^{***}$p$<$0.001.
\end{tablenotes}
\end{table}

\end{document}